# A Deep Learning Method for Real-time Bias Correction of Wind Field Forecasts in the Western North Pacific

Wei Zhang[a], Yueyue Jiang[a], Junyu Dong[*,a], Xiaojiang Song[*,b], Renbo Pang[b], Boyu Guan[b] and Hui Yu[c]

[a]*College of Computer Science and Technology, Ocean University of China, Qingdao, 266100, China*
[b]*National Marine Environment Forecasting Center, Beijing, 100082, China*
[c]*School of Creative Technologies, University of Portsmouth, Portsmouth, PO1 2DJ, UK*

### ABSTRACT

Forecasts by the European Centre for Medium-Range Weather Forecasts (ECMWF; EC for short) can provide a basis for the establishment of maritime-disaster warning systems, but they contain some systematic biases. The fifth-generation EC atmospheric reanalysis (ERA5) data have high accuracy, but are delayed by about 5 days. To overcome this issue, a spatiotemporal deep-learning method could be used for nonlinear mapping between EC and ERA5 data, which would improve the quality of EC wind forecast data in real time. In this study, we developed the Multi-Task-Double Encoder Trajectory Gated Recurrent Unit (MT-DETrajGRU) model, which uses an improved "double-encoder forecaster" architecture to model the spatiotemporal sequence of the U and V components of the wind field; we designed a multi-task learning loss function to correct wind speed and wind direction simultaneously using only one model. The study area was the western North Pacific (WNP), and real-time rolling bias corrections were made for 10-day wind-field forecasts released by the EC between December 2020 and November 2021, divided into four seasons. Compared with the original EC forecasts, after correction using the MT-DETrajGRU model the wind speed and wind direction biases in the four seasons were reduced by 8–11% and 9–14%, respectively. In addition, the proposed method modelled the data uniformly under different weather conditions. The correction performance under normal and typhoon conditions was comparable, indicating that the data-driven mode constructed here is robust and generalizable.

## 1. Introduction

The western North Pacific (WNP) is the north-western part of the Pacific Ocean. It is located adjacent to the Asian continent in the west and is connected to the Indian and Arctic oceans in the north and south, respectively. Theocean currents flowing through this area include the Kuroshio, warm North Pacific, and cold Kuril currents. Due to its geographical location and various ocean currents, the WNP has become one of the most important sea areas for tropical cyclone (TC) formation Magee et al. (2021), and about 30% of the world's high-strength TCs have occurred in this area Woo and Park (2021). TCs are among the most destructive natural disasters Matsuura et al. (2003), and are typically associated with strong winds and rainy weather. The wind field also drives large waves, and strong winds cause large fluctuations in the sea surface, including storm surges Li and Wang (2016). This not only threatens the property and safety of residents in coastal areas, but also affects maritime transportation. Therefore, accurate forecasting of sea surface wind fields in the WNP is essential for timely measures to mitigate natural disasters.

Numerical weather prediction (NWP) models, such as the European Centre for Medium-Range Weather Forecasts (ECMWF; EC for short) are widely used for weather forecasting and disaster prevention and mitigation Shen et al. (2020). EC models are based on physical equations, but due to the limitation of model resolution, physical parameters are used to resolve sub grid scale processes Wang et al. (2021). The uncertainty of the physical parameters and model structure; therefore results in, systematic biases in forecasts to some extent (Xu et al. (2021); Laloyaux et al. (2022)). Therefore, to improve the forecast accuracy of NWP models, an effective post-processing method of bias correction is needed.

[*]Corresponding author
dongjunyu@ouc.edu.cn(J. Dong); xjsong@nmefc.cn(X. Song)
ORCID(s): 0000-0001-7390-7613(W. Zhang); 0000-0003-1451-4865(Y. Jiang)

In recent decades, many post-processing methods for bias correction of NWP models have been proposed. The traditional statistical models that are widely used in wind field forecasting and bias correction include the perfect prognostic (PP) Klein et al. (1959), model output statistics (MOS) Glahn and Lowry (1972) and autoregressive moving average (ARMA) Torres et al. (2005) models. The PP model uses a statistical screening procedure to extract relevant variables from a large number of variables with which to construct linear regression equations to complete correction of the NWP model, the MOS model implements correction by building linear statistical models of observations with numerical forecast estimates of a set of related atmospheric variables; these two methods provide the basis for improving forecasting ability. Meanwhile, one study adjusted the parameters of the f-ARIMA model to identify long-term wind speed patterns and predict wind speeds during the next 1–2 days Kavasseri and Seetharaman (2009). Another proposed the "anomaly numerical correction with observations" (ANO) method Peng et al. (2013), which integrates historical ERA-interim reanalysis data, to forecast severe weather and heavy precipitation. However, due to the linear assumptions of traditional statistical models, they are unable to perform nonlinear fitting, and it is difficult for them to deal with nonlinear wind field data Duan et al. (2021).

Several types of neural network have been developed, including the convolutional neural network (CNN) Zhu et al. (2017), long short-term memory (LSTM) network Liu et al. (2018), deep belief network (DBN) Wan et al. (2016) and autoencoder (AE) network Mezaache and Bouzgou (2018), etc. These models have strong self-learning ability and can effectively fit nonlinear data Liu et al. (2020). Therefore, neural-network models have recently been used for the prediction and bias correction of marine variables. In addition, a very important type of deep learning is multi-task learning Ruder (2017), in which one model can complete multiple tasks simultaneously through sharing of the bottom layer or use of a multi-task learning loss function Duong et al. (2015). Multi-task models can ensure model accuracy and integrate multiple tasks into a single model, which makes the training and maintenance of the model more convenient. A dropout technique with a probabilistic neural network (D-PNN) has been applied to correct short-term wind speed forecast bias Zjavka (2015), and a sequence transfer correction algorithm (STCA) was proposed to correct wind speeds based on the dynamic transfer relationship between wind speeds at times t and t + 1 Wang et al. (2019). The gradient matrix of wind speed has been used as a behavioural feature to build a 3D-CNN for predictions Zhu et al. (2021), and a gated recurrent unit (GRU) neural network has been applied to construct a weighted time series to correct wind speeds Ding et al. (2019). CU-net, which considers wind direction data as scalar data and inputs EC and fifth generation ECMWF reanalysis (ERA5) data for bias correction of gridded wind direction forecasts has also been proposed Han et al. (2021). One study used a one-dimensional (1D)-CNN to model a 1D time series, to predict wind speed and wind direction Harbola and Coors (2019), and another modelled a non-stationary time series of wind directions considering cyclic features Solari and Losada (2016).

The influence of the cyclic characteristics of wind direction vectors cannot be avoided in any of the models described above. Therefore, a more reasonable modelling method to improve the accuracy of wind speed and wind direction correction is needed. Some studies showed that the limitations of wind direction forecasts can be overcome by evaluating the U and V components Turbelin et al. (2009) and that better results can be achieved by modelling the U and V components than by just using wind speed data Zhang et al. (2006). Inspired by these studies, we treated the wind field as a vector field and modelled its U and V components to correct wind speed and wind direction. Due to the dynamic and thermal relationships between different ocean variables, and because changes in the wind field depend on the temporal and spatial dimensions, we constructed a spatiotemporal sequence for each wind component. For medium-term forecasts (e.g. the EC wind field over for the next 10 days), accuracy is dependent on the initial conditions to some extent Cheng et al. (2017). The strength of the correlation between forecast and ground truth values gradually decreases with an increase in the forecast horizon Yang et al. (2022), which complicates improvement of the forecast accuracy of medium-term NWP models.

In this study, we established the Double Encoder Trajectory GRU (DETrajGRU) model for correction of wind speed or direction, and the Multi-Task DETrajGRU (MT-DETrajGRU) model for simultaneous correction of wind speed and wind direction. The MT-DETrajGRU model has the following key contributions. The challenge of the cyclic characteristics of wind direction bias correction is effectively solved by modelling the spatiotemporal sequence of the U and V components of surface wind vector components at 10 m height above sea level. A multi-task learning loss function was designed to complete wind speed and wind direction bias correction using only one model.

The remainder of this paper is structured as follows. Section 2 describes the study area and data used in the experiments. It also describes the framework of the wind field bias-correction model. Section 3 describes the sample-construction method, model structure, loss function, and model evaluation standards. Section 4 discusses the seasonal wind field correction results and compares the correction performance of the MT-DETrajGRU model under normal and typhoon conditions. Finally, we present the conclusions and discuss future research in Section 5.

**Table 1**
Strong and super typhoons occurring in December 2020 to November 2021.

| TC | time of activity | maximum intensities |
|---|---|---|
| Surigae | April 14, 2021 to April 24, 2021 | 220km/h |
| In-fa | July 18, 2021 to July 28, 2021 | 151km/h |
| Chanthu | September 07, 2021 to September 18, 2021 | 245km/h |
| Mindulle | September 23, 2021 to October 02, 2021 | 200km/h |
| Nyatoh | November 30, 2021 to December 04, 2021 | 198km/h |

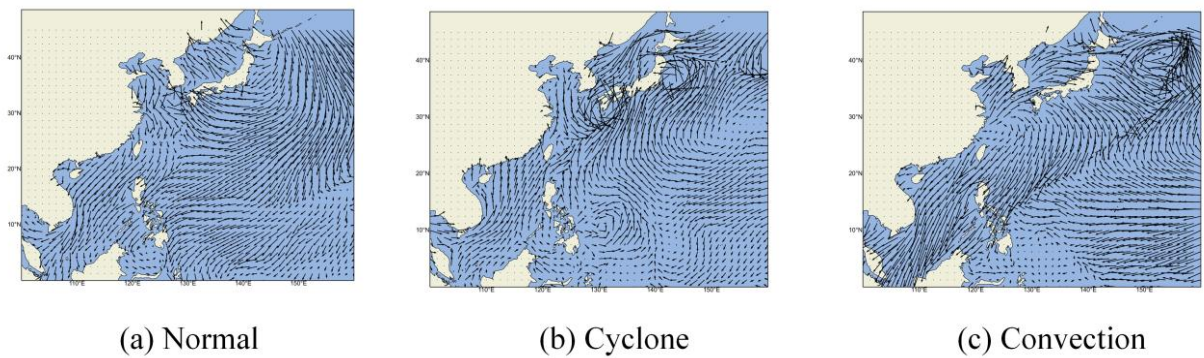

**Figure 1:** Vector diagram under (a) normal, (b) cyclonic and (c) convective weather conditions in the study area. The direction and length of each arrow represent the wind direction and wind speed, respectively.

## 2. Study area, data and proposed model
### 2.1. Study Area

The study area was between 0–45°N and 100–160°E (Figure 1), and a statistical analysis of the ERA5 sea and land wind field values in this area for the month of January 2021 is shown in Figure 2. Due to the influence of frictional forces caused by buildings and terrain factors (Ren et al. (2018); Wu et al. (2018)), land wind speeds were relatively low and land wind directions were significantly more unstable than sea wind directions; therefore, we masked the land area and studied only the sea surface wind field in the WNP. We plotted the ERA5 data for this area as a vector diagram (Figure 1) and found that cyclones and convection occurred in the region during more than half of the study period. The cyclone in Figure 1(b) occurred on August 8, 2021, and the convection in Figure 1(c) occurred on January 13, 2021. In Table 1, we list the strong and super typhoons that occurred during our study period (December 2020 to November 2021). Although EC forecasts of tropical cyclones are generally better than those of other global models, the forecasts of TC intensity still show a systematic negative bias. The WNP has a high frequency of typhoons throughout the year, and therefore correcting typhoon-forecast data bias is challenging Chan et al. (2021).

Meanwhile, because the intensity and structure of TCs have highly complex spatiotemporal characteristics, and given that the asymmetric distribution of convection leads to variation in TC intensity (Yang et al. (2016); Sun et al. 2 020)), bias correction is difficult; this highlights the practical significance of our research. The WNP covers the main monsoon regions in the world, i.e. the East Asian and WNP monsoon regions, and there are significant seasonal variations in WNP winds (Krishnamurthy (2018); Cao et al. (2021)). Therefore, we analysed the WNP sea surface wind field data according to season: spring included March, April and May; summer included June, July and August; autumn included September, October and November; and winter included December, January and February.

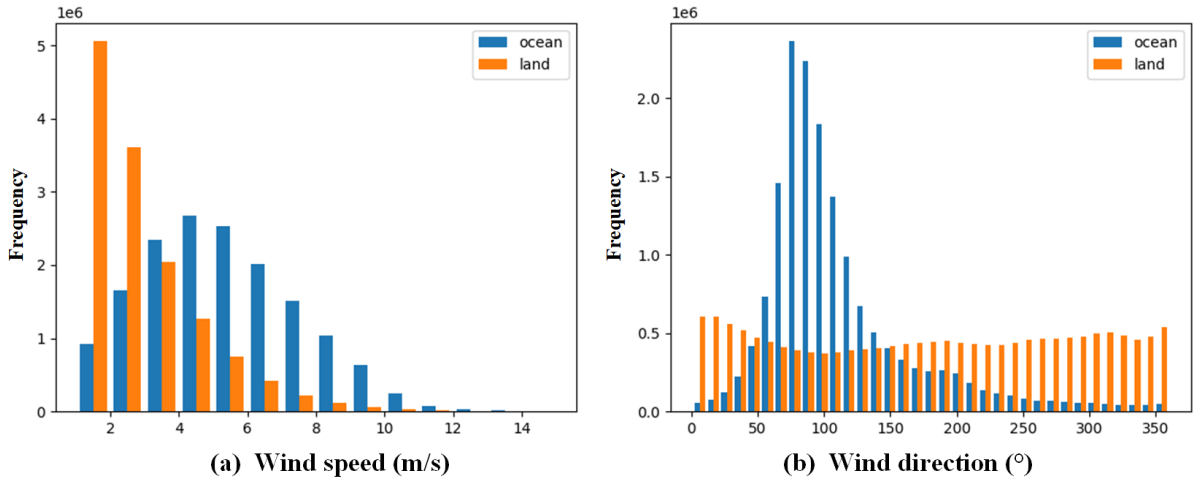

**Figure 2:** Statistical differences between the ocean and land wind fields. (a) wind speed, (b) wind direction.

### 2.2. Data

The wind forecast data used in this study were the grid forecast data released by the EC [1,2]. EC is a typical NWP model. According to the actual situation in the atmosphere, under certain initial values and boundary value conditions, a large-scale computer was used for numerical calculations to solve the motion state of the wind field over a certain period in the future Frnda et al. (2022). The spatial resolution of EC forecast data is 0.1°, and the forecast data for the next 10 days is updated at 00 and 12 UTC. The time resolution was 3 h for the first 6 days and 6 h for the last 4 days (total of 65 forecast times).

The ground truth is based on the ERA5 dataset [3], which uses advanced data assimilation techniques to incorporate as much observational data as possible, with high reliability, and is often used as the ground truth in studies of bias correction of numerical models (He et al. (2019); Hersbach et al. (2020); Han et al. (2021)). ERA5 can provide more accurate initial conditions for re-forecasts to obtain better forecast data [4]. Current EC forecasts are less accurate for certain weather phenomena, ERA5 can also provide guidance on where EC forecasting products are more/less accurate [5]. Therefore, we choose ERA5 data as the ground truth. It is updated daily with a delay of about 5 days and the spatial resolution is 0.25° and temporal resolution is 1 hour.

In this study, data were available from December 2020 to November 2021, based on the seasonal characteristics of wind data in the WNP Befort et al. (2018), we constructed correction models for spring, summer, autumn and winter, and the data for the four seasons were divided into training and testing sets. The 10-day forecast data at each "issue time" (every day at 00 and 12 UTC) were divided into three periods: days 1–2, 3–6, and 7–10. In this way, it was necessary to train only three models in each season to correct the wind field forecast data for the next 10 days.

### 2.3. Proposed model

An enhanced deep-learning model was constructed considering wind field as a vector field, and different loss functions were designed to correct wind speed and wind direction data. Wind speed and wind direction bias correction comprised four parts: data pre-processing, model improvement, training and testing (Figure 3).

First, after acquiring the EC and ERA5 data, the spatial resolution of the ERA5 data was increased from 0.25° to 0.1° by bilinear interpolation Han (2013). Spatiotemporal sequence samples corresponding to the U and V components were constructed using a "single-time-step rolling method" and divided into training and testing sets (see Section 3.1 for details).

---

[1] https://www.ecmwf.int/en/forecasts/datasets/set-i#I-i-a_fc
[2] https://apps.ecmwf.int/shopping-cart/orders/new/subset/162
[3] https://cds.climate.copernicus.eu/cdsapp#!/dataset/reanalysis-era5-single-levels?tab=overview
[4] https://www.ecmwf.int/en/newsletter/161/meteorology/use-era5-reanalysis-initialise-re-forecasts-proves-beneficial
[5] https://www.ecmwf.int/en/about/media-centre/science-blog/2017/era5-new-reanalysis-weather-and-climate-data

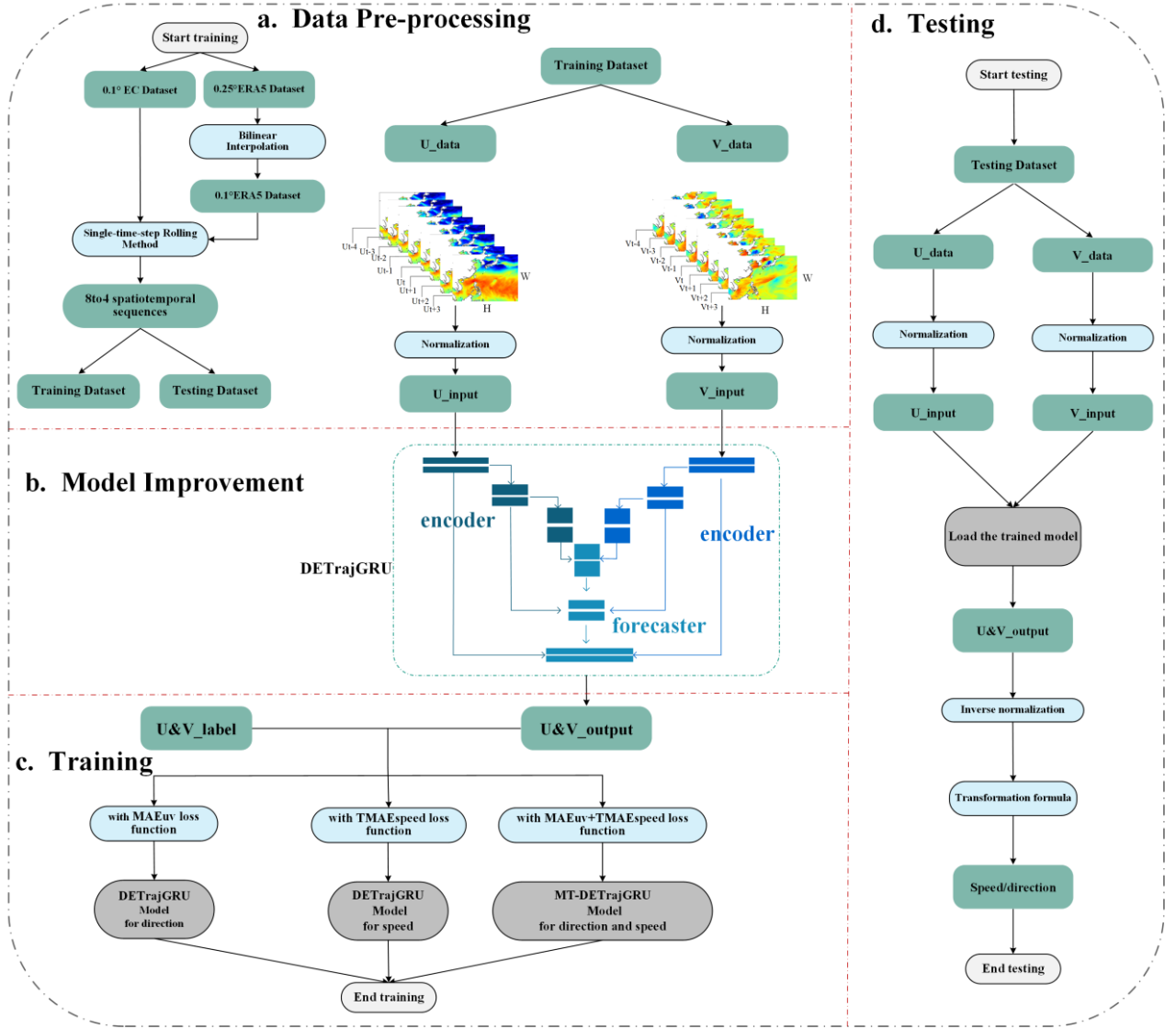

**Figure 3:** The framework of the proposed correction model.

Second, by improving the single-encoder forecaster structure of the TrajGRU model, the DETrajGRU model (based on the double-encoder forecaster) was constructed. The input data for this model were the U and V component samples. To extract the data for each wind component accurately, we used two encoders to build a dedicated "feature extraction path" for the U and V components. A forecaster was used to fuse the features and decode the results to obtain the model output. The details of the DETrajGRU network are discussed in Section 3.2.

Third, we trained the model using the training set. The classic L1 loss function (mean absolute error of the U and V components, $MAE_{UV}$) was used to train the model, and yielded the DETrajGRU model of wind direction. The transformed loss function (transformed mean absolute error of speed, $TMAE_{speed}$) was in accordance with the relationship between the U and V components and wind speed, and was used to obtain the DETrajGRU model of wind speed. The multi-task learning loss function ($MAE_{UV} + TMAE_{speed}$) was used to train the model, and yielded the MT-DETrajGRU model of wind speed and wind direction.

Finally, we used the testing set to evaluate the performance of the different models; wind speed and wind direction correction results were obtained by transforming the model output for the U and V components. The details of the transformation formula are discussed in Section 3.1.

## 3. Methodology
### 3.1. Data pre-processing and sample construction

In this section, we explain how the wind field spatiotemporal sequence bias correction problem was transformed into a video frame prediction problem, which enabled us to predict K-frame images based on historical S-frame images. The data used in this study were all gridded products; for example, the WNP wind forecast data at time t can be regarded as an M × N grid $x_t \in R^{H \times W}$. Each grid point can be regarded as a pixel point in each frame of the video, and the grid data of "S adjacent" forecast times can be regarded as constituting an S-frame image sequence expressed as a three-dimensional tensor, i.e. $X = \{x_1, x_2, \ldots, x_s\}, X \in R^{S \times H \times W}$. We constructed a many-to-many model for spatiotemporal samples with time sequence length S = 8 as the input and time sequence length K = 4 as the output, based on the U and V components, respectively.

The EC provides wind field forecast data for the next 10 days. As described in Section 2.2, at each issue time this comprises a total of 65 forecast times for the next 10 days, at 3hr interval for the first 6 days and 6hr interval for the last 4 days. We used the single-time-step rolling method to construct samples, which not only expanded the number of samples but also one model was sufficient to correct multiple forecast moments simultaneously. The 10-day forecast data were divided into three periods, as also delineated previously, such that the model did not have to be trained separately for each forecast time.

The single-time-step rolling method is shown in Figure 4. The time window was set to 8, and the samples were constructed by rolling backward one forecast moment at a time. For current time t, eight adjacent times were selected to form an input sequence $X = \{x_{t-4}, x_{t-3}, x_{t-2}, x_{t-1}, x_t, x_{t+1}, x_{t+2}, x_{t+3}\}, X \in R^{8 \times H \times W}$. The four most probable adjacent times comprised the output sequence $\hat{X} = \{\hat{x}_t, \hat{x}_{t+1}, \hat{x}_{t+2}, \hat{x}_{t+3}\}, \hat{X} \in R^{4 \times H \times W}$. The corresponding ground truth value of ERA5 was $Y = \{Y_t, Y_{t+1}, Y_{t+2}, Y_{t+3}\}, Y \in R^{4 \times H \times W}$, where H and W are the height and width of the input image, respectively (both = 240 in this study). Because our study area was a 450 × 600 grid and the input images were 240 × 240 in size, we formed a large image from six small images to cover the entire study area and test the model. The "weighted average fusion method" Song et al. (2009) was used to smooth overlapping areas and obtain the final correction result. Because ERA5 is updated daily (delayed by about 5 days), to construct samples using this method only the EC forecast data were used as input (Figure 4), which can ensure that as long as the EC issues the 10-day forecast, we can load the trained models and the corrected results could be obtained within 140 seconds on average.

It was necessary to normalize the wind components' samples before inputting them into the neural network model. Because the activation function used by the model for up- and down-sampling was the leaky rectified linear unit (LeakyReLU) Lu et al. (2021), we normalized the values to [-1,1]. We analysed the U and V component data and determined maximum and minimum are 25 m/s and -25 m/s, respectively. Data beyond this range were treated as the maxima and minima values. The normalization formula was as follows:

$$X_{normalization} = \frac{X - MIN}{MAX - MIN} * 2 - 1 \quad (1)$$

The model output was "inversely normalized"; formula 1 was used to obtain the actual value of the U and V components, and the final wind speed and wind direction were calculated using the transformation formula:

$$speed = f_1(U, V) = \sqrt{U^2 + V^2} \quad (2)$$

$$dir = f_2(U, V) = \begin{cases} 270 - \arctan(\frac{V}{U}) * 180/\pi & U > 0 \\ 90 - \arctan(\frac{V}{U}) * 180/\pi & U < 0 \\ 180 & U = 0 \text{ and } V > 0 \\ 360 & U = 0 \text{ and } V < 0 \\ 0 & U = 0 \text{ and } V = 0 \end{cases} \quad (3)$$

### 3.2. DETrajGRU model

In this section, we explain why we used the TrajGRU model, and then provide a detailed summary of the proposed DETrajGRU model.

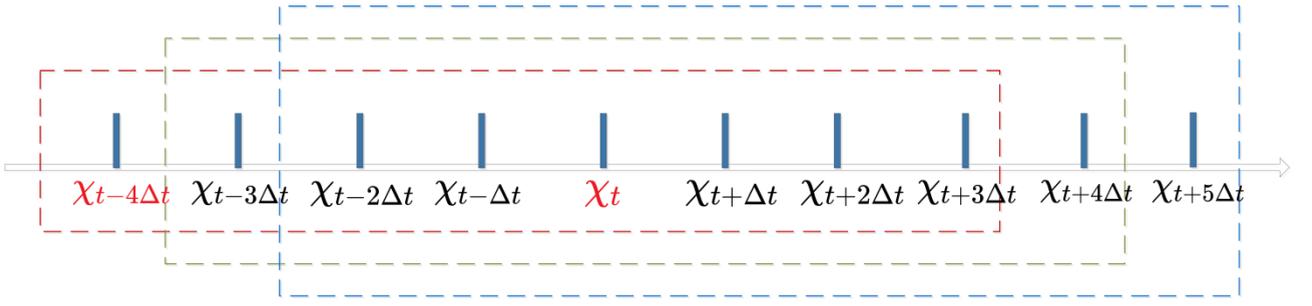

**Figure 4:** Schematic diagram of the single-time-step rolling method. Each blue column represents a 240 × 240 grid, the red text denotes the issue time, and the black text denotes the forecast time. For input sequence $X$ in each sliding window, $\Delta t$ = 3 h or 6 h.

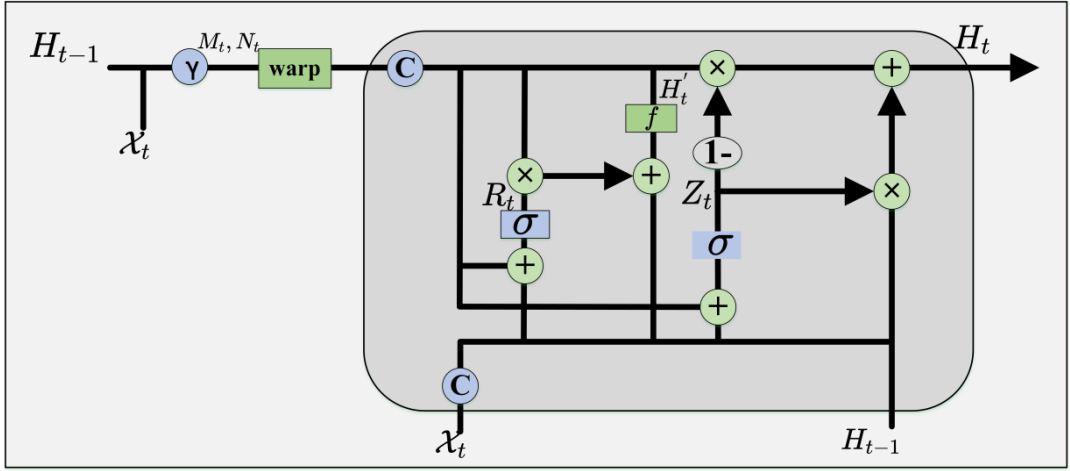

**Figure 5:** Internal structure of the TrajGRU layer, where C is the location-variant convolution, and × is the Hadamard product.

### 3.2.1. From FC-LSTM to ConvLSTM to TrajGRU

Fully connected long short-term memory (FC-LSTM) networks can effectively process 1D time sequences using contextual information, but cannot handle multi-dimensional spatiotemporal sequences. Therefore, classical convolution was used instead of the "full connection operation" to build a convolution LSTM (ConvLSTM) model to predict spatiotemporal sequences Shi et al. (2015). However, the recursive structure of the ConvLSTM model is location-invariant, whereas the natural movement and transformation wind field is usually location-variant. It is inefficient to use "location-invariant convolution" to learn "location-variant structures"; therefore, we used the TrajGRU model Shi et al. (2017) with a location-variant convolution that can dynamically determine the evolutionary trajectory of the wind field at each position. At each current time t and each position, according to input $x_t$ and hidden state $H_{t-1}$, flow fields storing the local connection structure $M_t$, $N_t$, were generated, and trajectory changes in the wind field were learned so that the points in the convolution operation were the most relevant with respect to the current position.

The internal structure of the TrajGRU layer is shown in Figure 5. After obtaining input $x_t$ for current time and hidden state $H_{t-1}$ in the previous moment, six important variables were used to update TrajGRU: new information $H'_t$, memory state $H_t$, reset gate $R_t$, update gate $Z_t$, the flow fields that store the local connection structure $M_t$, $N_t$. Here, $H_t, R_t, Z_t, H'_t \in R^{C_h \times H \times W}$, $x_t \in R^{C_i \times H \times W}$, where H and W are the height and width, and $C_h$, $C_i$ are the channel sizes of the hidden state and input tensors, respectively. Each variable was calculated at time t by the following formula:

$$M_t, N_t = \gamma(x_t, H_{t-1}) \qquad (4)$$

$$Z_t = \sigma(W_{xz} * x_t + \sum_{l=1}^{L} W_{hz}^l * warp(H_{t-1}, M_{t,l}, N_{t,l})) \quad (5)$$

$$R_t = \sigma(W_{xr} * x_t + \sum_{l=1}^{L} W_{hr}^l * warp(H_{t-1}, M_{t,l}, N_{t,l})) \quad (6)$$

$$H'_t = f(W_{xh} * x_t + R_t \times (\sum_{l=1}^{L} W_{hh}^l * warp(H_{t-1}, M_{t,l}, N_{t,l})))) \quad (7)$$

$$H_t = (1 - Z_t) \times H'_t + Z_t \times H_{t-1} \quad (8)$$

where $M_t, N_t \in R^{L \times H \times W}$ are generated by the structure-generating network $\gamma$, L is the total number of local links at each position, '*' represents convolution operation, '×' is the Hadamard product, $\delta$ is the sigmoid activation function, and $f(\cdot)$ is the LeakyReLU (negative slope set to 1 in this study). $W_{hz}^l, W_{hr}^l, W_{hh}^l$ are the weights used in the training process, and the $warp(\cdot)$ function selects positions identified by $M_{t,l}, N_{t,l}$ from $H_{t-1}$.

### 3.2.2 DETrajGRU model

Our goal was to build a deep learning model that can efficiently process samples based on the spatiotemporal sequences of the U and V components. The TrajGRU model based on the single-encoder forecaster structure can handle spatiotemporal sequences, but has only one encoder to extract features. When we entered multiple variables, the extracted features were mixed. Therefore, we introduced an encoder branch based on TrajGRU to obtain a "double-encoder forecaster structure". This can be used to create a dedicated feature extraction path for each wind component, so that the model can mine hidden wind field data for a single variable during feature map encoding. The feature maps were then fused by the forecaster to correct the results. The structure of the DETrajGRU model is shown in Figure 6.

In the encoding stage, we set the channel size of the two encoders to 1 and applied them separately to input sequences $\langle U_{t-4}, U_{t-3}, \ldots, U_{t+3} \rangle$ and $\langle V_{t-4}, V_{t-3}, \ldots, V_{t+3} \rangle$ to extract wind field information for each wind component. The forecaster was then used to fuse the information in the decoding stage and correct the forecast data; this yielded $\langle \widehat{U}_t, \widehat{U}_{t+1}, \widehat{U}_{t+2}, \widehat{U}_{t+3} \rangle$ and $\langle \widehat{V}_t, \widehat{V}_{t+1}, \widehat{V}_{t+2}, \widehat{V}_{t+3} \rangle$. In the DETrajGRU model, each encoder stacked three layers (downsampling layer + TrajGRU layer) to generate three hidden states. The hidden states of the U and V components after encoding were $U\_H^1, U\_H^2, U\_H^3 = h(U_{t-4}, U_{t-3}, \ldots, U_{t+3})$ and $V\_H^1, V\_H^2, V\_H^3 = h(V_{t-4}, V_{t-3}, \ldots, V_{t+3})$, where $U\_H^i, V\_H^i, \in R^{Ch \times H \times W}$. Then the $U\_H^i$ and $V\_H^i$ of each block were concatenated in the channel dimension to obtain hidden states $\langle H^1, H^2, \ldots, H^n \rangle$, where $H^i = [U\_H^i, V\_H^i], H^i \in R^{2Ch \times H \times W}$. Finally, using the forecaster, we obtained a three-layer block (upsampling layer + TrajGRU layer) to correct the results as follows: $\widehat{O}_t, \widehat{O}_{t+1}, \widehat{O}_{t+2}, \widehat{O}_{t+3}, = g(H^1, H^2, \ldots, H^n)$, where $\widehat{O}_i = [\widehat{U}_i, \widehat{V}_i], \widehat{O}_i \in R^{2 \times H \times W}$.

The stacking order of the three-layer blocks created by the encoder and forecaster was reversed, and feature maps were directly transferred from one block of the encoder to the corresponding block of the forecaster. In this way, the global spatiotemporal information contained within the high-order state could guide updating of the low-order state, thereby improving the accuracy of the corrected results.

### 3.3. Design of loss functions to train different models
### 3.3.1 Loss function

Correction of wind field forecast bias is a regression task, so all of the models in this study were regression models. The loss function of the regression model measured the gap between the model output and ground truth (ERA5). Our aim was to create a loss function to narrow the gap as much as possible. We designed loss functions for the different learning targets to construct bias correction models. When both the input and learning target was wind speed, $MAE_{speed}$ was used as the loss function. When both the input and learning target was wind direction, $MAEd_{dir}$ was used as the loss function (the reason for using MAEd are described in Section 3.4). When the input was the U and V components and the learning target was the wind direction, $MAE_{UV}$ was used as the loss function. In particular, when the input was the U and V components and the learning target was wind speed, because the relationship between the U and V components and wind speed was not linear, we designed a transformed $MAE_{speed}$ ($TMAE_{speed}$) loss function according to formula 2. When the input was the U and V components and the learning target was wind speed and wind direction, $MAE_{UV} + TMAE_{speed}$ was used as the multi-task learning loss function to train the model and correct wind speed and wind direction simultaneously. The loss functions were defined as follows:

$$MAE_{speed} = \frac{1}{L}\sum_{n=1}^{N}\sum_{i=1}^{H}\sum_{j=1}^{W}|speed_{n,i,j} - \widehat{speed}_{n,i,j}| \tag{9}$$

$$MAEd_{dir} = \frac{1}{L}\sum_{n=1}^{N}\sum_{i=1}^{H}\sum_{j=1}^{W}|E_{n,i,j}| \tag{10}$$

$$E_{n,i,j}\begin{cases} \widehat{dir}_{n,i,j} - dir_{n,i,j} & -180 \leq \widehat{dir}_{n,i,j} - dir_{n,i,j} \leq 180 \\ \widehat{dir}_{n,i,j} - dir_{n,i,j} + 360 & \widehat{dir}_{n,i,j} - dir_{n,i,j} < -180 \\ \widehat{dir}_{n,i,j} - dir_{n,i,j} - 360 & \widehat{dir}_{n,i,j} - dir_{n,i,j} > 180 \end{cases} \tag{11}$$

$$MAE_{UV} = \frac{1}{2} \times \frac{1}{L}\sum_{n=1}^{N}\sum_{i=1}^{H}\sum_{j=1}^{W}\left(|U_{n,i,j} - \widehat{U}_{n,i,j}| + |V_{n,i,j} - \widehat{V}_{n,i,j}|\right) \tag{12}$$

$$TMAE_{speed} = \frac{1}{L}\sum_{n=1}^{N}\sum_{i=1}^{H}\sum_{j=1}^{W}\left|\sqrt{U_{n,i,j}^2 + V_{n,i,j}^2} - \sqrt{\widehat{U}_{n,i,j}^2 + \widehat{V}_{n,i,j}^2}\right| \tag{13}$$

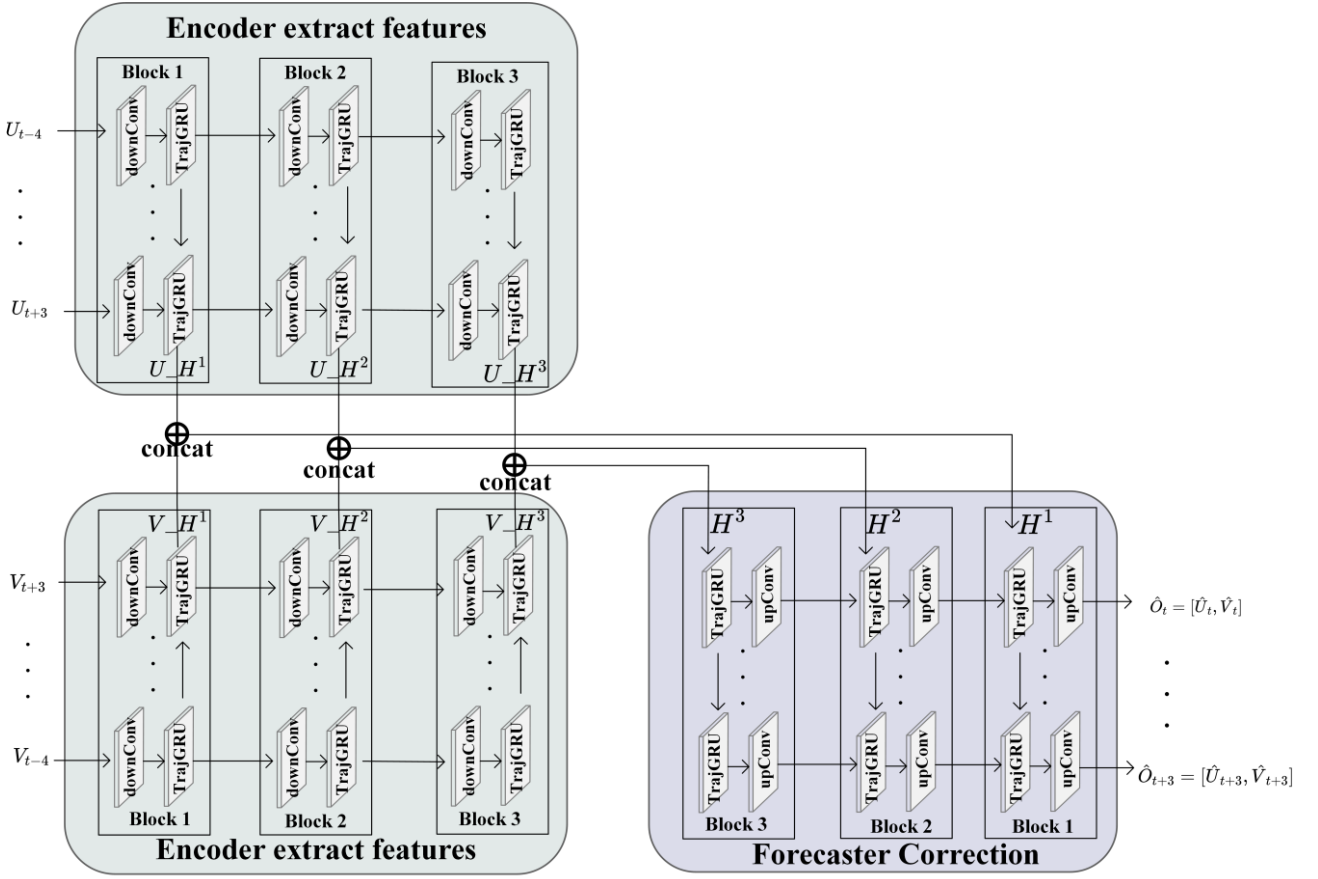

**Figure 6:** Structure of the DETrajGRU model.

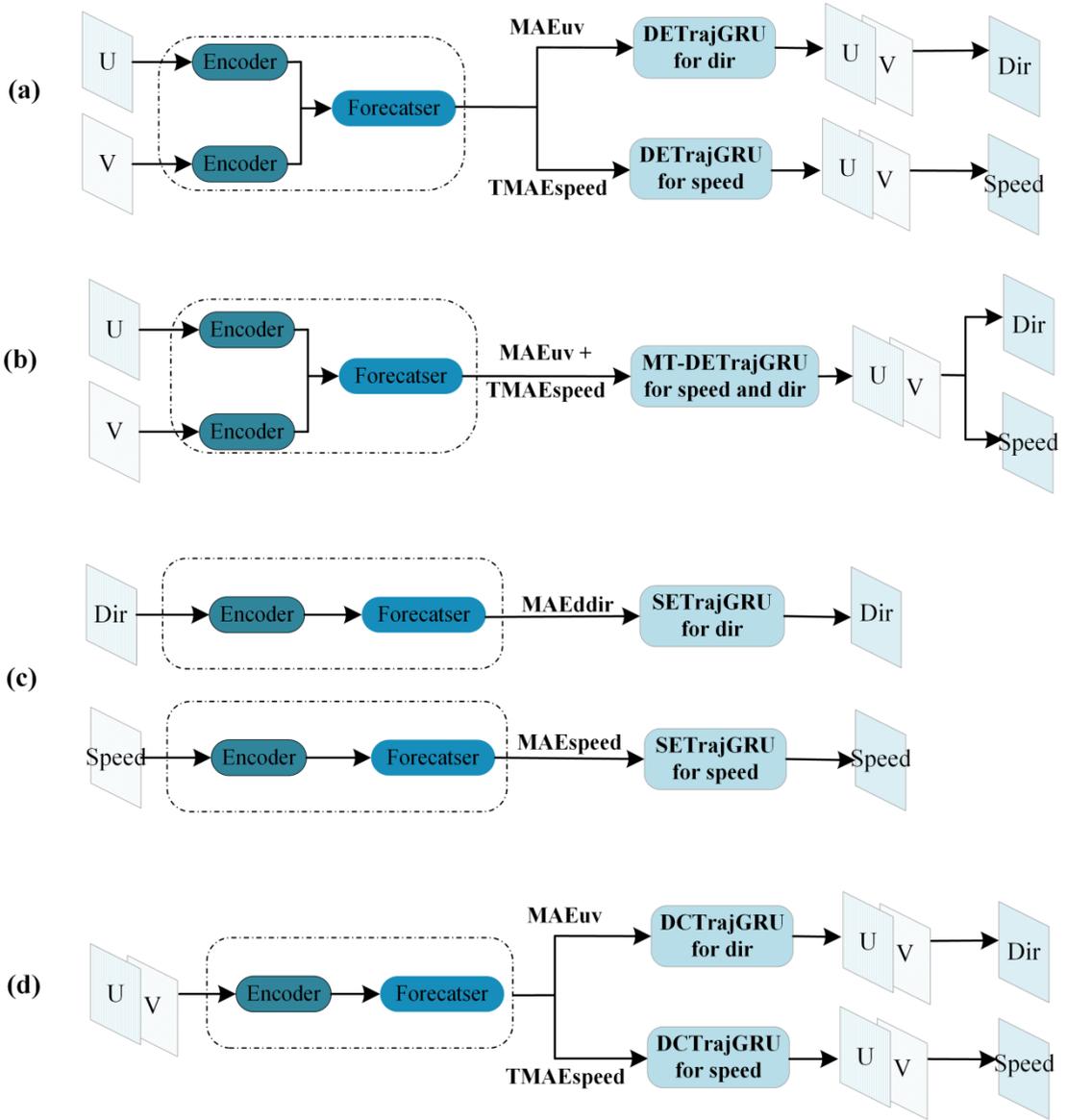

**Figure 7:** Flowchart of the correction models: (a) DETrajGRU, (b) MT-DETrajGRU, (c) SETrajGRU and (d) DCTrajGRU.

where $L = N \times H \times W, N = B \times T$, B is the number of samples in each batch, T is the time sequence length of output, in this study, B = 4, T = 4. H and W are the height and width of the model output, respectively, here, H=W=240. $\hat{\ast}_{n,i,j}$ is the model output value and $\ast_{n,i,j}$ is the the ground truth, where * stands for speed, dir, U or V.

### 3.3.2. The flowchart of different models

Figure 7 presents a flowchart of the correction models. Figure 7a shows the DETrajGRU model with the double-encoder forecaster structure, which used the U and V components as inputs and considered the combined effect of meridional and zonal winds on the evolution of the wind field. However, the model could correct only wind speed or wind direction. Figure 7b shows the MT-DETrajGRU model, which also used the U and V components as inputs but simultaneously corrected wind speed and wind direction. Figure 7c shows the SETrajGRU model, which served as the baseline model in this study. The SETrajGRU model was based on the single-encoder forecaster structure and used either wind speed or wind direction data as input. Figure 7d shows the Double-channel TrajGRU (DCTrajGRU) model, which was also based on the single-encoder forecaster structure but used the U and V components as the inputs. To enable the DCTrajGRU model to extract features from two wind components simultaneously using one encoder, we set the input channel of the encoder to 2.

**Table 2**
Configuration information of the wind direction correction models.

| model | input | output | learning target | channel | loss function |
|---|---|---|---|---|---|
| DETrajGRU | U and V | U and V | dir | 1 | $MAE_{UV}$ |
| MT-DETrajGRU | U and V | U and V | speed and dir | 1 | $MAE_{UV} + TMAE_{speed}$ |
| SETrajGRU | dir | dir | dir | 1 | $MAEd_{dir}$ |
| DCTrajGRU | U and V | U and V | dir | 2 | $MAE_{UV}$ |

We demonstrated that the MT-DETrajGRU model could correct two variables at the same time by comparing models (a) and (b) (Figure 7). We also demonstrated the effectiveness of the double-encoder forecaster structure by comparing models (a) and (d), and demonstrated the effectiveness of modelling the U and V components by comparing models (c) and (d).

### 3.4. Performance evaluation criteria

To evaluate model performance intuitively, we used two evaluation metrics. The most commonly used evaluation index ($MAE$) for regression models was used to evaluate model performance in terms of wind speed, defined as follows:

$$MAE = \frac{1}{T*H*W}\sum_{t=1}^{T}\sum_{i=1}^{H}\sum_{j=1}^{W}|y_{t,i,j} - \hat{y}_{t,i,j}| \qquad (14)$$

where N is the number of test samples, and H and W are the height and width of the study area, respectively. Here, H=450, W=600.

Regarding the wind direction, 0° and 360° point in the same direction, even though their numerical values are very different; thus, we must consider the cyclic characteristics of wind direction. According to the "QXT 229–2014 Verification Method for Wind Forecast" Han et al. (2021), $MAE$d represents a difference between the corrected and ground truth values of < 180°, and is appropriate for wind direction assessment. Therefore, this metric was used for our evaluation; it is defined as follows:

$$MAEd = \frac{1}{T*H*W}\sum_{t=1}^{T}\sum_{i=1}^{H}\sum_{j=1}^{W}|E_{t,i,j}| \qquad (15)$$

where $E_{t,i,j}$ is defined in formula 11.

## 4. Experiments and results
### 4.1. Corrected wind direction results for the four seasons

The details of the MT-DETrajGRU, DETrajGRU, SETrajGRU and DCTrajGRU wind direction correction models are shown in Table 2. Figure 8 shows the changes in $MAEd$ over the four seasons according to the different models (for the data, refer to "Appendix.pdf" file). The 10-day forecast data were divided into three periods, as explained above; therefore, only three models were needed per season for wind direction correction.

The DETrajGRU model of wind direction achieved the smallest MAEd for all four seasons, and exhibited better correction performance compared to the SETrajGRU and DCTrajGRU models. The wind direction correction performance of the MT-DETrajGRU model was comparable to that of the DETrajGRU model. Compared with the original EC 10-day forecast, the MAEd decreased by 9–14% when using the MT-DETrajGRU model. The histogram in Figure 8 shows that the MAEd for the EC wind direction forecast in the third period (days 7–10) was > 30 °. The wind direction input range was [0 °, 360 °] for the SETrajGRU model, i.e. the data fluctuated substantially and exhibited a distribution very different from the ground truth. Therefore, the corrected forecast was less accurate than the original EC forecast. However, the MT-DETrajGRU model showed good performance with respect to the U and V components. The wind components range was narrow [-25 m/s, 25 m/s] and the influence of the cyclic characteristics of the wind direction vector was avoided to a certain extent. The MAEd was reduced by about 4% by the MT-DETrajGRU model compared with the original EC forecast. Therefore, the U and V components were modelled more accurately.

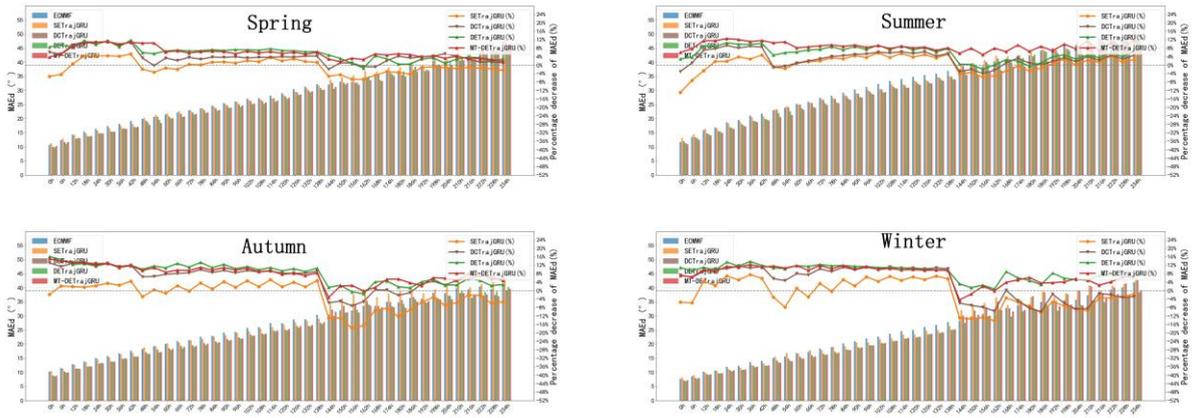

**Figure 8:** Changes in the MAEd after correcting for bias in wind direction forecasts for the four seasons using the MT-DETrajGRU, DETrajGRU, SETrajGRU and DCTrajGRU models (note: to display the correction results for 10 days in one image, the time resolution for the first 6 days, shown as 6h in the figure, was actually 3h in the correction).

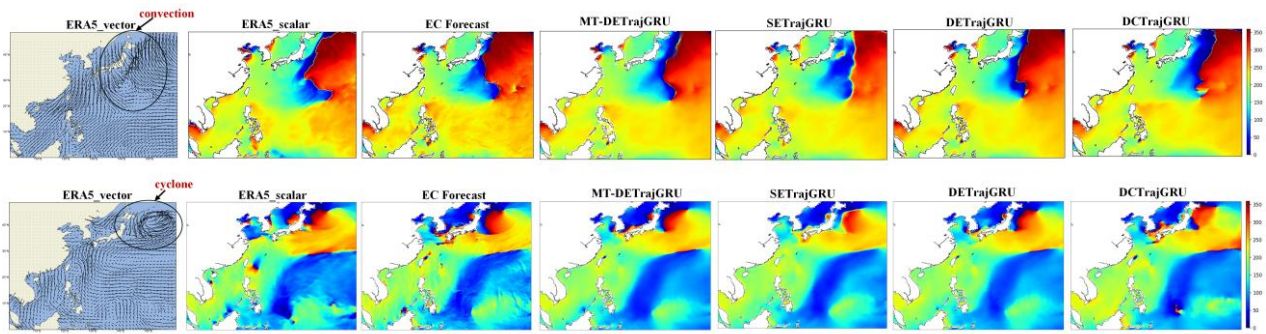

**Figure 9:** Examples of the corrected results of different models under convective and cyclonic conditions.

In order to accurately compare the correction performance of different models, examples of wind direction correction results are plotted (Figure 9). For the convection area, we plotted the corrected results at 00 UTC on February 02, 2021. The wind direction distributions of the MT-DETrajGRU, DETrajGRU and DCTrajGRU models in the convection area were more accurate and reasonable, and the contour line of the convection area was closer to ERA5 values after using the MT-DETrajGRU correction. For the cyclone area, we plotted the corrected results at 00 UTC on August 11, 2021. Although the corrected results of the MT-DETrajGRU model were close to the original EC forecast distribution, the MAEd before correction was 17.6 °, and that after correction was 15.1 °. These two examples show that the MT-DETrajGRU model can effectively correct the forecast bias of the wind direction vector under abnormal weather conditions.

The loss function used by the DETrajGRU model for wind direction was $MAE_{UV}$, which is not ideal for correcting wind speed. Formula 2 shows that the relationship between the U and V components and wind speed is not linear.

**Table 3**
Configuration information of the wind speed correction models.

| model | input | output | learning target | channel | loss function |
|---|---|---|---|---|---|
| DETrajGRU | U and V | U and V | speed | 1 | $TMAE_{speed}$ |
| MT-DETrajGRU | U and V | U and V | speed and dir | 1 | $MAE_{UV} + TMAE_{speed}$ |
| SETrajGRU | speed | speed | speed | 1 | $MAE_{speed}$ |
| DCTrajGRU | U and V | U and V | speed | 2 | $TMAE_{speed}$ |

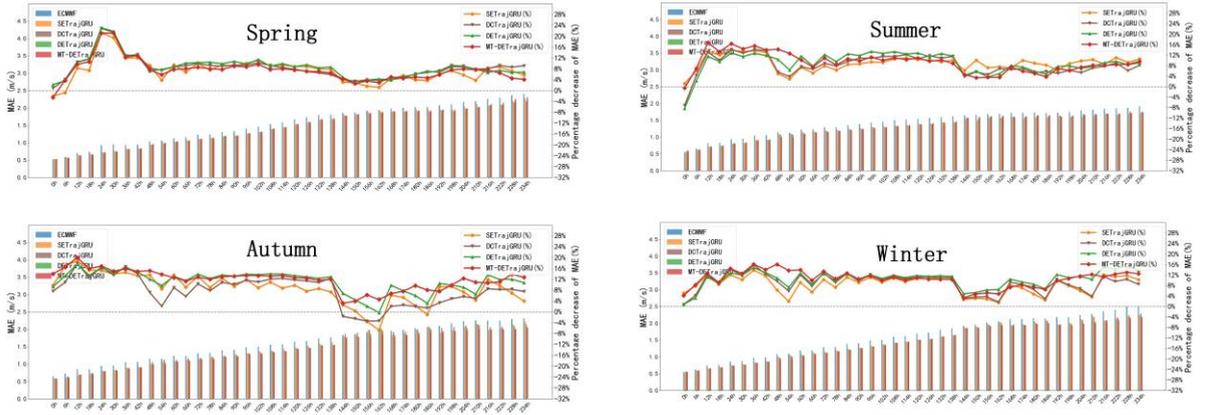

**Figure 10:** Same as Figure 8, but for the MAE of wind speed.

Therefore, for wind speed, the loss function was transformed ($TMAE_{speed}$) to train the DETrajGRU model for wind speed forecast (refer to Section 4.2 for details).

### 4.2. Corrected wind speed results for the four seasons

We trained the DETrajGRU model for wind speed forecast using the transformed loss function ($TMAE_{speed}$). Also, the wind speed correction performance of this model was compared with that of the MT-DETrajGRU model using $MAE_{UV} + TMAE_{speed}$ as the loss function. Wind speed correction also used the SETrajGRU model as the baseline, and the DCTrajGRU model was used to verify the effectiveness of the double-encoder forecaster architecture (Table 3).

Figure 10 shows the changes in MAE over the four seasons according to the different models (for the data, refer to "Appendix.pdf" file). Regarding wind speed, the correction performance of the DETrajGRU and MT-DETrajGRU models was comparable. Compared with the original EC forecast, the MAE decreased by 8–11% when using the MT-DETrajGRU model for 10-day wind speed forecasting. The MT-DETrajGRU model achieved the best correction results overall. Especially the third period (days 7–10) in autumn and winter, the correction accuracy of the MT-DETrajGRU model was about 2–5% higher than that of the SETrajGRU model.

The above correction results for wind speed and wind direction that biases therein could be resolved using a model with a multi-task learning loss function. The correction performance of the MT-DETrajGRU model was comparable to that of the DETrajGRU model only for a single learning target, indicating that the MT-DETrajGRU model has high generalizability and robustness. The U and V components could be modelled to correct wind speed and wind direction, which avoided the problem caused by the cyclical characteristics of the wind direction. Moreover, the DETrajGRU model outperformed the DCTrajGRU model, indicating the effectiveness of the proposed double-encoder forecaster structure. Notably, the bias statistics were based on the average of all grid points in the study area, and reflected the model's ability to correct the data for the entire study area reliably.

### 4.3. Scatterplot of 10-day corrected results

To compare the correction performance of the MT-DETrajGRU and SETrajGRU models for all grid points in the study area for 10-day forecasts, we produced a scatterplot of the model-corrected and ERA5 (ground truth) data (Figure 11). The abscissa represents the value of the ERA5 data, and the ordinate represents the EC forecast or model-corrected result. In a scatterplot, data distributed closer to the diagonal line denote a stronger linear relationship between two variables Kim et al. (2021). For presentation purposes, we constructed a scatterplot every 24 h for the corrected results; the issue time was 12 UTC on November 16, 2021. The scatter plots between the EC forecast and ERA5 data (red box in Figure 11) shows that with an increase in forecast horizon, the data distribution became more diffuse, and the linear correlation gradually weakened.

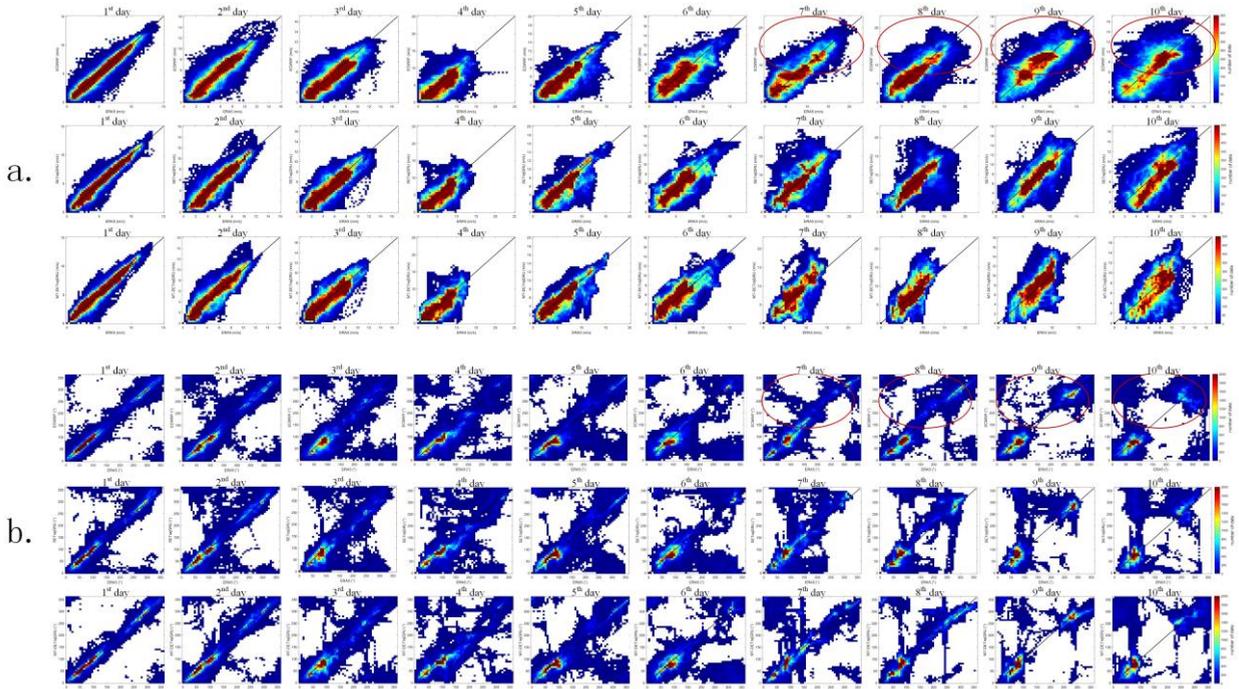

**Figure 11:** Scatter plots between the ERA5 data and corrected (a) wind speed and (b) wind direction data for a 10-day forecast. For each set of graphs, the EC forecasts, SETrajGRU correction results, and MT-DETrajGRU correction results are shown in the top, middle and bottom rows, respectively.

For wind speed (Figure 11a), the corrected data of both the MT-DETrajGRU and SETrajGRU models were closer to the diagonal than the original EC forecast data. The distorted shape of the MT-DETrajGRU model data distribution was significantly reduced after correction, such that the linear relationship was stronger than for the SETrajGRU model. In the third period (days 7–10), the EC wind speed forecast exhibited large bias (range: 8–13 m/s; red box in Figure 11a). This was reduced by correction using the MT-DETrajGRU model, such that the results showed a better fit with the ERA5 data.

For wind direction (Figure 11b) was overestimated by the EC (range: 50–200 °; red box in Figure 11b), especially in the third period. The data distribution after correction by the SETrajGRU model was more distorted than that of the original EC forecast, whereas distortion was significantly reduced after correction using the MT-DETrajGRU model (i.e. the data were closer to the diagonal). This shows that the U and V components allow for more accurate wind direction forecasts, by avoiding influence of the cyclical characteristics of wind direction during bias correction.

### 4.4. Case study

To verify the bias correction performance of the MT-DETrajGRU model under normal and typhoon conditions, example wind speed and wind direction correction results are presented. We subjected normal and typhoon condition test samples to independent statistical analyses. Because tropical storm Dujuan was active in the WNP from February 17 to February 22, 2021, this was selected as the typhoon period. For comparison, samples from February 7, 2021 to February 12, 2021 were used to represent normal weather conditions. The MAE and MAEd values before and after wind speed and wind direction correction obtained at 00 UTC every day under the two types of weather conditions are shown in Table 4, with an increase in forecast horizon, the bias of the original EC forecast gradually increased. When the forecast horizon was 6 days, the original forecast wind speed and direction

**Table 4**

Statistical results for wind speed and wind direction under normal and typhoon weather conditions.

| Variable | Metric | Condition | | Correction Horizon (days) | | | | | |
|---|---|---|---|---|---|---|---|---|---|
| | | | | 1 | 2 | 3 | 4 | 5 | 6 |
| Wind speed | MAE(m/s) | normal | before | 0.56 | 0.89 | 0.92 | 1.17 | 1.51 | 1.89 |
| | | | after | 0.53 | 0.72 | 0.73 | 1.00 | 1.41 | 1.71 |
| | | typhoon | before | 0.66 | 0.92 | 1.04 | 1.12 | 1.56 | 1.64 |
| | | | after | 0.58 | 0.79 | 0.86 | 0.93 | 1.39 | 1.45 |
| Wind direction | MAEd(°) | normal | before | 9.21 | 11.34 | 11.13 | 17.26 | 24.23 | 28.42 |
| | | | after | 8.50 | 9.42 | 9.57 | 15.19 | 22.14 | 25.84 |
| | | typhoon | before | 8.85 | 9.69 | 10.92 | 15.67 | 20.62 | 23.92 |
| | | | after | 7.75 | 8.74 | 9.50 | 13.85 | 19.00 | 20.33 |

biases under typhoon conditions were 1.64 m/s and 23.92 °, respectively, after correction using the MT-DETrajGRU model the biases were reduced to 1.45 m/s and 20.33 °, respectively. Under normal and typhoon conditions, the average percentage decreases in MAE at all times after wind speed correction were 12.6% and 13.8%, respectively, while the average percentage decreases in MAEd at all times after wind direction correction were 11.4% and 11.6%, respectively; thus, MT-DETrajGRU model correction performance was satisfactorily under both normal and typhoon conditions.

The distributions of the correction results for the MT-DETrajGRU model under normal and typhoon conditions are shown in Figure 12. In terms of wind direction (Figure 12 a,b), for the convective area, the contours of the correction results under the two weather conditions were highly similar to the ERA5 data. For the cyclone area, the similarity of the correction results to the ERA5 data was lower than that for the convective area, but the cyclone's locations were predicted accurately. This may have been because the changes in wind direction in the cyclone area exhibited a spiral shape, and were more unstable than the changes in wind direction in the convective area Yoshida et al. (2008). In terms of wind speed (Figure 12 c, d), although the corrected wind speed of the MT-DETrajGRU model was slightly smaller than that of ERA5, the correction results were very similar to the ERA5 data in terms of the overall contours.

Overall, the MT-DETrajGRU model exhibited comparable performance in terms of wind speed and wind direction corrections under typhoon and normal weather conditions, indicating that the model showed good generalizability, and it is not necessary to divide and train the model according to different weather conditions.

## 5. Conclusions

Our study showed that a data-driven deep-learning method can simultaneously correct wind speed and wind direction forecast biases at the spatiotemporal scale. This method uses multi-task learning and an improved DETrajGRU model, i.e. the MT-DETrajGRU model. By modelling the spatiotemporal sequence of the U and V components, bias correction for wind speed and wind direction can be achieved using only one model. The double encoder forecaster structure can provide the feature extraction and feature fusion path for the U and V components, respectively. Real-time bias correction can be achieved; when the EC updates forecasts, the model can correct 10-day forecasts in a few seconds.

Model performance was evaluated in terms of the correction of WNP wind speed and wind direction data. Considering the dynamic balance between wind components and seasonal variation in WNP wind field data, we used EC wind field forecast data and ERA5 reanalysis data from December 2020 to November 2021 for model training and testing. We used the MAE and MAEd as indicators of model performance. First, MT-DETrajGRU-corrected results were compared to the original EC forecasts. The MAE of wind speed decreased by 8–11% for the 10- day forecast, while the MAEd of wind direction decreased by 9–14%. The correction performance of the MT-DETrajGRU and DETrajGRU models was comparable according to the experiments, indicating that wind speed and wind direction can be corrected simultaneously, reducing model training time and hardware requirements by almost half. Finally, tests of independent samples obtained under normal and typhoon conditions showed comparable wind speed and wind direction correction performance for the MT-DETrajGRU model under different weather conditions, indicating good model generalizability and robustness to different weather conditions.

In subsequent research, we will also use temperature, atmospheric pressure, humidity, and other variables that affect the wind field as model inputs. It can be seen from the experimental results that strong winds are weakened after wind speed correction using the proposed model, and areas of strong and weak wind speeds can be corrected respectively. This study provides a basis for further wind field prediction research, especially that using deep-learning techniques to improve the forecast accuracy of NWP models.

## Acknowledgment

This paper was supported by National Natural Science Foundation (grant 42276202 and U1706218), and the National Key Research and Development Program of China (grant 2018YFC1407001).

## References

Befort, D.J., Wild, S., Knight, J.R., Lockwood, J.F., Thornton, H.E., Hermanson, L., Bett, P.E., Weisheimer, A., Leckebusch, G.C., 2018. Seasonal forecast skill for extratropical cyclones and windstorms. Quarterly Journal of the Royal Meteorological Society doi:10.1002/qj.3406.

Cao, X., Wu, R., Xu, J., Feng, J., Zhang, X., Dai, Y., Liu, Y., 2021. Contribution of the intensity of intraseasonal oscillation to the interannual variation of tropical cyclogenesis over the western north pacific. Environmental Research Communications 3, 031002.

Chan, M.H.K., Wong, W.K., Au-Yeung, K.C., 2021. Machine learning in calibrating tropical cyclone intensity forecast of ecmwf eps. Meteorological applications 28, e2041.

Cheng, W.Y., Liu, Y., Bourgeois, A.J., Wu, Y., Haupt, S.E., 2017. Short-term wind forecast of a data assimilation/weather forecasting system with wind turbine anemometer measurement assimilation. Renewable Energy 107, 340–351.

Ding, M., Zhou, H., Xie, H., Wu, M., Nakanishi, Y., Yokoyama, R., 2019. A gated recurrent unit neural networks based wind speed error correction model for short-term wind power forecasting. Neurocomputing 365, 54–61.

Duan, J., Zuo, H., Bai, Y., Duan, J., Chang, M., Chen, B., 2021. Short-term wind speed forecasting using recurrent neural networks with error correction. Energy 217, 119397. doi:10.1016/j.energy.2020.119397.

Duong, L., Cohn, T., Bird, S., Cook, P., 2015. Low resource dependency parsing: Cross-lingual parameter sharing in a neural network parser, in: Proceedings of the 53rd annual meeting of the Association for Computational Linguistics and the 7th international joint conference on natural language processing (volume 2: short papers), pp. 845–850.

Frnda, J., Durica, M., Rozhon, J., Vojtekova, M., Nedoma, J., Martinek, R., 2022. Ecmwf short-term prediction accuracy improvement by deep learning. Scientific Reports 12, 1–11.

Glahn, H.R., Lowry, D.A., 1972. The use of model output statistics (mos) in objective weather forecasting. Journal of Applied Meteorology and Climatology 11, 1203–1211.

Han, D., 2013. Comparison of commonly used image interpolation methods, in: Conference of the 2nd International Conference on Computer Science and Electronics Engineering (ICCSEE 2013), Atlantis Press. pp. 1556–1559.

Han, L., Chen, M., Chen, K., Chen, H., Zhang, Y., Lu, B., Song, L., Qin, R., 2021. A deep learning method for bias correction of ecmwf 24–240 h forecasts. Advances in Atmospheric Sciences 38, 1444–1459.

Harbola, S., Coors, V., 2019. One dimensional convolutional neural network architectures for wind prediction. Energy Conversion and Management 195, 70–75. doi:10.1016/j.enconman.2019.05.007.

He, D., Zhou, Z., Kang, Z., Liu, L., 2019. Numerical studies on forecast error correction of grapes model with variational approach. Advances in Meteorology 2019.

Hersbach, H., Bell, B., Berrisford, P., Hirahara, S., Horányi, A., Muñoz-Sabater, J., Nicolas, J., Peubey, C., Radu, R., Schepers, D., 2020. The era5 global reanalysis. Quarterly Journal of the Royal Meteorological Society 146, 1999–2049.

Kavasseri, R.G., Seetharaman, K., 2009. Day-ahead wind speed forecasting using f-arima models. Renewable Energy 34, 1388–1393.

Kim, D.K., Suezawa, T., Mega, T., Kikuchi, H., Yoshikawa, E., Baron, P., Ushio, T., 2021. Improving precipitation nowcasting using a three-dimensional convolutional neural network model from multi parameter phased array weather radar observations. Atmospheric Research 262, 105774.

Klein, W.H., Lewis, B.M., Enger, I., 1959. Objective prediction of five-day mean temperatures during winter. Journal of Atmospheric Sciences 16, 672–682.

Krishnamurthy, V., 2018. Intraseasonal oscillations in east asian and south asian monsoons. Climate Dynamics 51, 4185–4205.

Laloyaux, P., Kurth, T., Dueben, P.D., Hall, D., 2022. Deep learning to estimate model biases in an operational nwp assimilation system. Journal of Advances in Modeling Earth Systems, e2022MS003016.

Li, Z., Wang, H., 2016. Ocean wave simulation based on wind field. Plos one 11, e0147123.

Liu, H., Mi, X., Li, Y., 2018. Smart multi-step deep learning model for wind speed forecasting based on variational mode decomposition, singular spectrum analysis, lstm network and elm. Energy Conversion and Management 159, 54–64.

Liu, Z., Jiang, P., Zhang, L., Niu, X., 2020. A combined forecasting model for time series: Application to short-term wind speed forecasting. Applied Energy 259, 114137.

Lu, T., Chen, T., Gao, F., Sun, B., Ntziachristos, V., Li, J., 2021. Lv-gan: A deep learning approach for limited-view optoacoustic imaging based on hybrid datasets. Journal of biophotonics 14, e202000325.

Magee, A.D., Kiem, A.S., Chan, J.C., 2021. A new approach for location-specific seasonal outlooks of typhoon and super typhoon frequency across the western north pacific region. Scientific reports 11, 1–16.

Matsuura, T., Yumoto, M., Iizuka, S., 2003. A mechanism of interdecadal variability of tropical cyclone activity over the western north pacific. Climate Dynamics 21, 105–117. doi:10.1007/s00382-003-0327-3.

Mezaache, H., Bouzgou, H., 2018. Auto-encoder with neural networks for wind speed forecasting, in: 2018 International Conference on Communications and Electrical Engineering (ICCEE), IEEE. pp. 1–5.

Peng, X., Che, Y., Chang, J., 2013. A novel approach to improve numerical weather prediction skills by using anomaly integration and historical data. Journal of Geophysical Research: Atmospheres 118, 8814–8826.

Ren, H., Laima, S., Chen, W.L., Zhang, B., Guo, A., Li, H., 2018. Numerical simulation and prediction of spatial wind field under complex terrain. Journal of Wind Engineering and Industrial Aerodynamics 180, 49–65.

Ruder, S., 2017. An overview of multi-task learning in deep neural networks. arXiv preprint arXiv:1706.05098 .

Shen, X., Wang, J., Li, Z., Chen, D., Gong, J., 2020. Research and operational development of numerical weather prediction in china. Journal of Meteorological Research 34, 675–698. doi:10.1007/s13351-020-9847-6.

Shi, X., Chen, Z., Wang, H., Yeung, D.Y., Wong, W.K., Woo, W.c., 2015. Convolutional lstm network: A machine learning approach for precipitation nowcasting. Advances in neural information processing systems 28.

Shi, X., Gao, Z., Lausen, L., Wang, H., Yeung, D.Y., Wong, W.k., Woo, W.c., 2017. Deep learning for precipitation nowcasting: A benchmark and a new model. Advances in neural information processing systems 30.

Solari, S., Losada, M.Á., 2016. Simulation of non-stationary wind speed and direction time series. Journal of Wind Engineering and Industrial Aerodynamics 149, 48–58.

Song, L., Lin, Y., Feng, W., Zhao, M., 2009. A novel automatic weighted image fusion algorithm, in: 2009 International Workshop on Intelligent Systems and Applications, IEEE. pp. 1–4.

Sun, C., Liu, Y., Gong, Z., Kucharski, F., Li, J., Wang, Q., Li, X., 2020. The footprint of atlantic multidecadal oscillation on the intensity of tropical cyclones over the western north pacific. Frontiers in Earth Science , 565.

Torres, J.L., Garcia, A., De Blas, M., De Francisco, A., 2005. Forecast of hourly average wind speed with arma models in navarre (spain). Solar energy 79, 65–77.

Turbelin, G., Ngae, P., Grignon, M., 2009. Wavelet cross-correlation analysis of wind speed series generated by ann based models. Renewable Energy 34, 1024–1032.

Wan, J., Liu, J., Ren, G., Guo, Y., Yu, D., Hu, Q., 2016. Day-ahead prediction of wind speed with deep feature learning. International Journal of Pattern Recognition and Artificial Intelligence 30, 1650011.

Wang, H., Han, S., Liu, Y., Yan, J., Li, L., 2019. Sequence transfer correction algorithm for numerical weather prediction wind speed and its application in a wind power forecasting system. Applied Energy 237, 1–10.

Wang, Y., Zou, R., Liu, F., Zhang, L., Liu, Q., 2021. A review of wind speed and wind power forecasting with deep neural networks. Applied Energy 304, 117766. doi:10.1016/j.apenergy.2021.117766.

Woo, H.J., Park, K.A., 2021. Estimation of extreme significant wave height in the northwest pacific using satellite altimeter data focused on typhoons (1992–2016). Remote Sensing 13, 1063. doi:10.3390/rs13061063.

Wu, J., Zha, J., Zhao, D., Yang, Q., 2018. Changes in terrestrial near-surface wind speed and their possible causes: an overview. Climate dynamics 51, 2039–2078.

Xu, W., Liu, P., Cheng, L., Zhou, Y., Xia, Q., Gong, Y., Liu, Y., 2021. Multi-step wind speed prediction by combining a wrf simulation and an error correction strategy. Renewable Energy 163, 772–782.

Yang, D., Wang, W., Hong, T., 2022. A historical weather forecast dataset from the european centre for medium-range weather forecasts (ecmwf) for energy forecasting. Solar Energy 232, 263–274.

Yang, L., Fei, J., Huang, X., Cheng, X., Yang, X., Ding, J., Shi, W., 2016. Asymmetric distribution of convection in tropical cyclones over the western north pacific ocean. Advances in Atmospheric Sciences 33, 1306–1321.

Yoshida, M., Yamamoto, M., Takagi, K., Ohkuma, T., 2008. Prediction of typhoon wind by level 2.5 closure model. Journal of wind engineering and industrial aerodynamics 96, 2104–2120.

Zhang, Y., Liu, P., Pun, B., Seigneur, C., 2006. A comprehensive performance evaluation of mm5-cmaq for the summer 1999 southern oxidants study episode—part i: Evaluation protocols, databases, and meteorological predictions. Atmospheric Environment 40, 4825–4838.

Zhu, A., Li, X., Mo, Z., Wu, R., 2017. Wind power prediction based on a convolutional neural network, in: 2017 International Conference on Circuits, Devices and Systems (ICCDS), IEEE. pp. 131–135.

Zhu, X., Liu, R., Chen, Y., Gao, X., Wang, Y., Xu, Z., 2021. Wind speed behaviors feather analysis and its utilization on wind speed prediction using 3d-cnn. Energy 236, 121523.

Zjavka, L., 2015. Wind speed forecast correction models using polynomial neural networks. Renewable Energy 83, 998–1006.

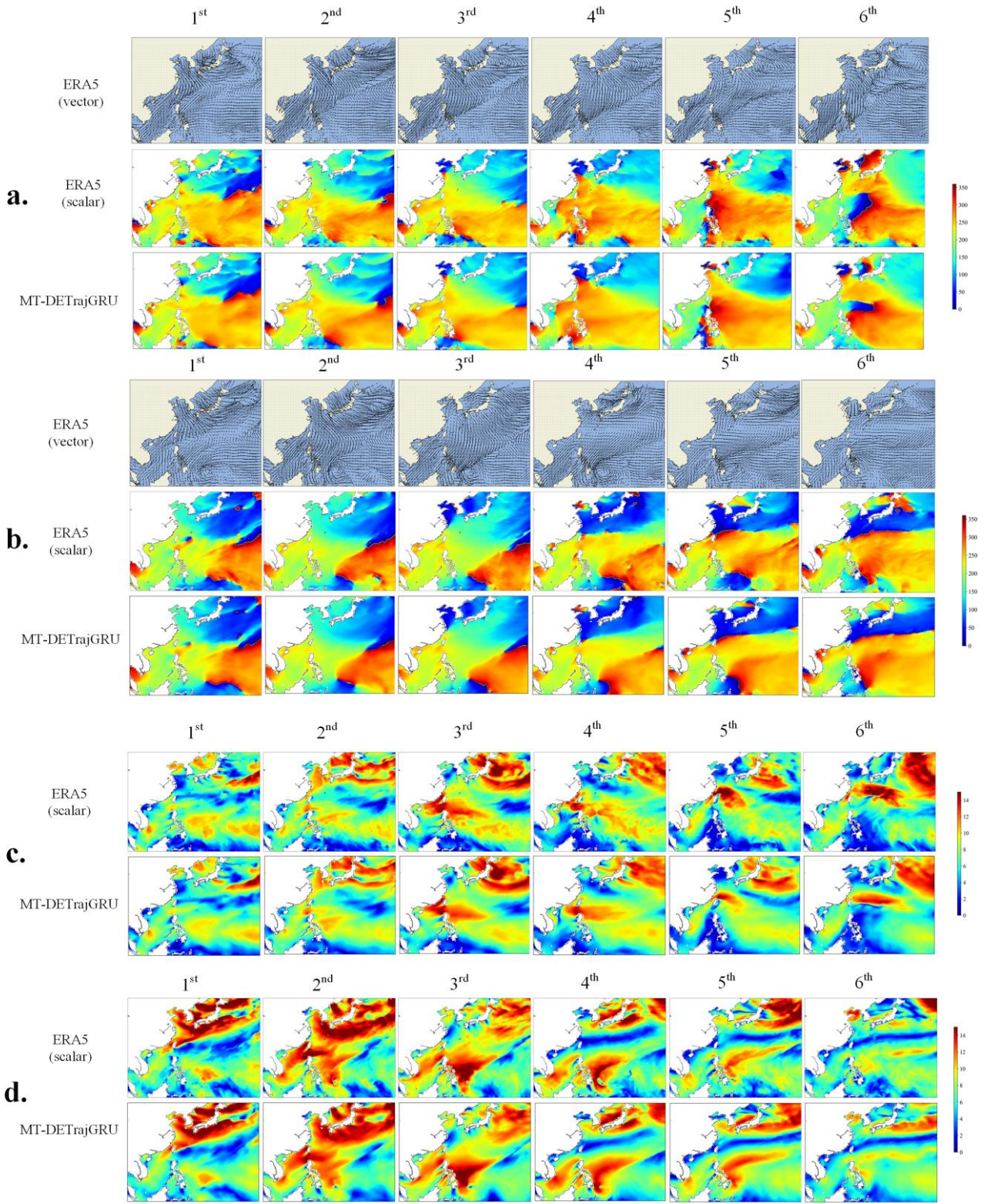

**Figure 12:** Comparison of the wind speed and wind direction correction results of the MT-DETrajGRU model under normal and typhoon conditions. (a and b) Wind direction results under normal and typhoon weather conditions, respectively. (c and d) Wind speed results under normal and typhoon conditions, respectively.